\documentclass[aps,prl,twocolumn,showpacs,superscriptaddress,showkeys,bm,amsmath,amssymb]{revtex4-1}
\usepackage{graphicx}
\usepackage{xcolor}

\begin{document}
\title{Measurement of Reactor Antineutrino Flux and Spectrum at RENO}
\newcommand{\CNU}{\affiliation{Institute for Universe and Elementary Particles, Chonnam National University, Gwangju 61186, Korea}}
\newcommand{\DSU}{\affiliation{Institute for High Energy Physics, Dongshin University, Naju 58245, Korea}}
\newcommand{\GIST}{\affiliation{GIST College, Gwangju Institute of Science and Technology, Gwangju 61005, Korea}}
\newcommand{\IBS}{\affiliation{Institute for Basic Science, Daejeon 34047, Korea}}
\newcommand{\KAIST}{\affiliation{Department of Physics, Korea Advanced Institute of Science and Technology, Daejeon 34141, Korea}}
\newcommand{\KNU}{\affiliation{Department of Physics, Kyungpook National University, Daegu 41566, Korea}}
\newcommand{\SNU}{\affiliation{Department of Physics and Astronomy, Seoul National University, Seoul 08826, Korea}}
\newcommand{\SYU}{\affiliation{Department of Fire Safety, Seoyeong University, Gwangju 61268, Korea}}
\newcommand{\SKKU}{\affiliation{Department of Physics, Sungkyunkwan University, Suwon 16419, Korea}}

\author{S.~G. Yoon}\KAIST
\author{H.~Seo}\SNU
\author{Z.~Atif}\CNU
\author{J.~H.~Choi}\DSU
\author{H.~I.~Jang}\SYU
\author{J.~S.~Jang}\GIST
\author{S.~H.~Jeon}\SKKU
\author{K.~K.~Joo}\CNU
\author{K.~Ju}\KAIST
\author{D.~E.~Jung}\SKKU
\author{J.~G.~Kim}\SKKU
\author{J.~H.~Kim}\SKKU
\author{J.~Y.~Kim}\CNU
\author{S.~B.~Kim}\SKKU
\author{S.~Y.~Kim}\SNU
\author{W.~Kim}\KNU
\author{E.~Kwon}\SKKU
\author{D.~H.~Lee}\SKKU
\author{H.~G.~Lee}\SNU
\author{I.~T.~Lim}\CNU
\author{D.~H.~Moon}\CNU
\author{M.~Y.~Pac}\DSU
\author{J.~W.~Seo}\SKKU
\author{C.~D.~Shin}\CNU
\author{B.~S.~Yang}\SNU
\author{J.~Yoo}\SNU
\author{I.~S.~Yeo}\CNU
\author{I.~Yu}\SKKU

\collaboration{The RENO Collaboration}
\noaffiliation
\date{\today}
\begin{abstract}
{The RENO experiment reports measured flux and energy spectrum of reactor electron antineutrinos\,($\overline{\nu}_e$) from the six reactors at Hanbit Nuclear Power Plant. The measurements use 966\,094\,(116\,111)\,$\overline{\nu}_e$ candidate events with a background fraction of 2.39\%\,(5.13\%), acquired in the near\,(far) detector, from August 2011 to March 2020. The inverse beta decay (IBD) yield is measured as (5.852$\,\pm\,$0.094$) \times 10^{-43}$\,cm$^2$/fission, corresponding to 0.941\,$\pm$ 0.015 of the prediction by the Huber and Mueller (HM) model. A reactor $\overline{\nu}_e$ spectrum is obtained by unfolding a measured IBD prompt spectrum. The obtained neutrino spectrum shows a clear excess around 6\,MeV relative to the HM prediction. The obtained reactor $\overline{\nu}_e$ spectrum will be useful for understanding unknown neutrino properties and reactor models. The observed discrepancies suggest the next round of precision measurements and modification of the current reactor $\overline{\nu}_e$ models.}
\end{abstract}
\maketitle 
	
\par A fission reactor is an intense source of $\overline{\nu}_e$ produced in the beta decays of neutron-rich nuclei. Nuclear reactors have played crucial roles in the impressive progress of neutrino physics from neutrino discovery to recent oscillation results. The predicted rate and energy spectrum of the reactor $\overline{\nu}_e$ depend on the instantaneous thermal power and fission fraction of four dominant isotopes in the nuclear fuel, as well as on the details of their fission process involving thousands of short-lived isotopes. The Huber and Mueller\,(HM) model\,\cite{Huber2011,Mueller2011} predicts the reactor $\overline{\nu}_e$ rate and spectrum based on a conversion of measured fission $\beta$ spectra and \textit{ab initio} calculation. According to the HM prediction, there exists $\sim$5\% deficit in the observed reactor $\overline{\nu}_{e}$ rate, so-called reactor antineutrino anomaly\,\cite{RAA2011}. In addition, the discrepancy varies according to the fuel composition of the reactors providing the observed $\overline{\nu}_e$ rate\,\cite{Daya2017EvolutionFuel,RENO2018fueldependent}. The recent study finds that the systematic uncertainty related to the handling of the forbidden nuclear transitions in the calculation can be up to 4\% \cite{Hayes2014}. RENO\,\cite{RENO_5MeV_boston, RENO2018mee} and other reactor experiments\, \cite{DayaBay2016_5MeV,DoubleChooz2020Nature,NEOS2016prl,Goesgen,PROSPECT2020,STEREO2020} have observed an excess of events in the measured IBD prompt energy spectrum at 5\,MeV relative to the HM prediction. This observation suggests the needs for reevaluation and modification of the reactor $\overline{\nu}_e$ model as well as for precise measurements. This Letter reports RENO's first measurement of the reactor $\overline{\nu}_e$ flux and unfolded spectrum based on 966\,094\,(116\,111) IBD candidate events in the near\,(far) detector. This result provides useful information for unveiling anomalies associated with reactor neutrinos and unknown neutrino properties.
	
\par The RENO experiment consists of near and far detectors located at 294 and 1\,383\,m, respectively, from the center of the six reactor cores of the Hanbit Nuclear Power Plant, Yonggwang, Korea. The near (far) detector is under 120\,m (450\,m) of water equivalent overburden. Six pressurized water reactors, each with a maximum thermal output of 2.8\,GW$_{\text{th}}$, are situated in a linear array spanning 1.3\,km with equal spacing. The reactor flux-weighted baseline is 419.4\,m for the near detector and 1\,447.1\,m for the far detector. 
	
\par The reactor $\overline{\nu}_e$ is detected through IBD interaction, $\overline{\nu}_e + p\,\rightarrow\,e^+ + n$, with free protons in hydrocarbon liquid scintillator with 0.1\% gadolinium\,(Gd) as a target. The coincidence of a prompt positron signal and a $26\,\mu s$ delayed signal from neutron capture by Gd provides the distinctive IBD signature against backgrounds. The prompt signal releases the energy of 1.02\,MeV as two $\gamma$-rays coming from the electron-positron annihilation in addition to the positron kinetic energy. The delayed signal produces several $\gamma$-rays with the total energy of $\sim8$\,MeV. A detailed description of RENO experimental setup can be found in Ref. \cite{RENO2018mee}.
	
\par An IBD yield in a detector can be predicted by a reactor $\overline{\nu}_e$ flux and the IBD cross section. With the fairly well-known IBD cross section and the number of target protons, the reactor $\overline{\nu}_e$ flux can be measured from the number of reactor $\overline{\nu}_e$ events ($n_{\nu}$). The observed $n_{\nu}$ is given by,
\begin{equation}
n_{\nu}	=\,\overline{y}_f \sum^6_{r=1} \frac{N_p}{4 \pi L^2_r} \int \frac{W_{\text{th},\,r}(t)\overline{P}_r(t)}{\sum_i f_{i,\,r}(t) \overline{E}_i}\,\epsilon_d(t) dt\:,
\label{eq:expect_number_of_neutrino}
\end{equation}
\noindent where $N_p$ is the number of the target protons, $L_r$ is the distance between a detector and $r$-th reactor, $f_{i,\,r}(t)$ is the fission fraction for the $i$-th isotope in the $r$-th reactor, $\overline{E}_i$ is the average energy released per fission of $i$-th isotope, $W_{th,r}(t)$ is the thermal power of the $r$-th reactor, $\overline{P}_r(t)$ is the mean survival probability of $\overline{\nu}_e$ from the $r$-th reactor, $\epsilon_d(t)$ is the detection efficiency, and $\overline{y}_f$ is the IBD yield per fission averaged over the four main isotopes during the detector operating period.
	
\par Uncertainties of reactor fission fractions and thermal powers, provided by the power plant, contribute 0.7\% and 0.5\%, respectively, to the error in the total IBD yield of each reactor. These uncertainties are treated uncorrelated among the six reactors. The average effective fission fractions of $^{235}$U, $^{238}$U, $^{239}$Pu and $^{241}$Pu over the operating period are 0.571, 0.073, 0.300 and 0.056 for the near detector, and 0.574, 0.073, 0.298, and 0.055 for the far detector, respectively. The average energy released per fission is given in Ref.\,\cite{Ma2013_average_energy}.
	
\par Detection efficiency is estimated by using control samples and a Monte Carlo simulation\,(MC)\,\cite{RENO2018PRD}. The fractional uncertainty of the overall detection efficiency is 1.41\% and the largest source of the measured IBD yield error. For precision measurement of an absolute reactor $\overline{\nu}_e$ flux, the detection efficiency needs to be accurately determined. In this analysis, several improvements are made in the evaluation of detection efficiency components and their uncertainties compared to the previous ones\,\cite{RENO2018PRD}. The only updated efficiencies and uncertainties are briefly described below. The uncertainty of target protons is corrected from 0.5\% to 0.7\% after a detailed study of the hydrogen composition and the density of the Gd-doped liquid scintillator. The Gd capture fraction is changed from (85.48\,$\pm$\,0.48)\% to (84.95\,$\pm$\,0.80)\% by taking into account the neutron spill-out effect that was neglected previously. The spill-in efficiency is reevaluated with an improved method, using distributions of prompt vertex position and neutron capture time in data. Because the spill-in favors a reconstructed prompt vertex outside the neutrino target and thus a longer delayed-signal time, the distributions are effective tools to estimate its contribution to an IBD candidate sample. The spill-in efficiency is updated from (2.00\,$\pm$\,0.61)\% to (1.34\,$\pm$\,0.69)\% where the uncertainty of the new efficiency mostly comes from the difference between the data and MC. The efficiency of prompt energy requirement is changed from 98.77\% to 97.95\% according to the updated spill-in efficiency. The uncertainty of delayed energy requirement is changed from 0.50\% to 0.70\% based on an improved MC study of the spectral shape. As a result, the new detection efficiency is estimated to be (74.87\,$\pm$\,1.06)\%. Each component of detection efficiency and corresponding systematic uncertainty are summarized in Table~\ref{table:det_eff}. The correlated uncertainty between the near and far detectors is 1.05\%, much larger than the uncorrelated uncertainty of 0.14\%. The IBD signal loss due to the muon timing veto and requirements is also updated as $(40.0\,\pm\,0.01)$\% for the near detector and $(31.1\,\pm\,0.01)$\% for the far detector. 
	
\begin{table}[t!]
\caption{Cross section, number of target protons, and detection efficiencies of IBD selection criteria. The individual efficiencies are given below the total efficiency. The uncertainty includes both correlated and uncorrelated components between the near and far detectors.}

\begin{tabular}{lr}
\hline\hline
    IBD total cross section                & $\pm$ 0.09\%   \\
	Target protons                   & (1.189    $\pm$ 0.008)  $\times 10^{30}$                    \\
	Detection efficiency\,(total)    & (74.87   $\pm$           1.06)\%          \\ \hline
	Trigger efficiency               & 99.77    $\pm$         0.05           \\
	Qmax/Qtot                        & 100.00   $\pm$          0.02           \\
	Gd capture fraction              & 84.95    $\pm$          0.80           \\
	Spill-in                         & 101.34  $\pm$          0.69           \\ 
	Prompt energy requirement        & 97.95   $\pm$           0.10           \\
	Delayed energy requirement       & 92.14   $\pm$           0.70           \\
	Time coincidence                 & 96.59   $\pm$           0.26           \\
	Spatial correlation              & 100.00   $\pm$          0.03           \\ \hline\hline
\end{tabular}
\label{table:det_eff}
\end{table}
	
\par In order to measure the IBD yield with respect to the HM prediction, a $\chi^2$ minimization method is used. A ratio $R$ of observed IBD event rate relative to the HM prediction, together with the value of neutrino mixing angle $\theta_{13}$, is determined using a $\chi^2$ defined as,
\begin{align}
\chi^2  = & \sum_{d=N,\,F} \frac{\big[ O_d - R\,T_d \big]^2}{O_d+B_d} + \sum_{d=N,\,F} \Big( \frac{b^d}{\sigma^d_{\text{bkg}}} \Big)^2 \nonumber \\
   		  &  + \sum_{d=N,\,F} \Big( \frac{\xi^d_{\text{uncor}}}{\sigma_{\xi,\text{uncor}}} \Big)^2
		     + \Big( \frac{\xi_{\text{cor}}}{\sigma_{\xi,\text{cor}}} \Big)^2 + \sum_{r=1}^6 \Big( \frac{f_{\text{r}}}{\sigma_{f}} \Big)^2 ,  
\end{align}
\noindent where $O_d$ and $B_d$ are the numbers of the observed IBD and background events in $d$-th detector, respectively, $T_d=\sum_{r=1}^6 T^r_d (1+b^d+\xi^d_{\text{uncor}}+\xi_{\text{cor}}+f_{\text{r}})$ is the number of expected IBD events, $T^r_d$ is the number of expected IBD events from r-th reactor,  $\sigma^d_{\text{bkg}}$ is the background uncertainty, $\sigma_{\xi,\text{uncor}}$\,(0.19\%) and $\sigma_{\xi,\text{cor}}$\,(1.55\%) are the uncorrelated and correlated uncertainties of the detection efficiency, respectively, $\sigma_{f}$\,(0.9\%) is the reactor uncertainty correlated between the two detectors but uncorrelated among the six reactors, and $b^d$, $\xi^d_{\text{uncor}}$, $\xi_{\text{cor}}$ and $f_{\text{r}}$ are their corresponding pull parameters. Note that the correlated uncertainty of detection efficiency, 1.55\%, includes the uncertainties of the IBD cross section and the number of target protons. The best fit value of $R$ is determined by minimizing the $\chi^2$ and found to be 0.941\,$\pm$\,0.001\,(stat.)\,$\pm$ {0.015\,(sys.), reassuring the deficit of observed reactor $\overline{\nu}_e$ event rate relative to its prediction. Fig.~\ref{fig:flux_global} shows the reactor $\overline{\nu}_e$ rates measured at various distances from reactors.

\begin{figure}[t!]
\includegraphics[width=0.49\textwidth]{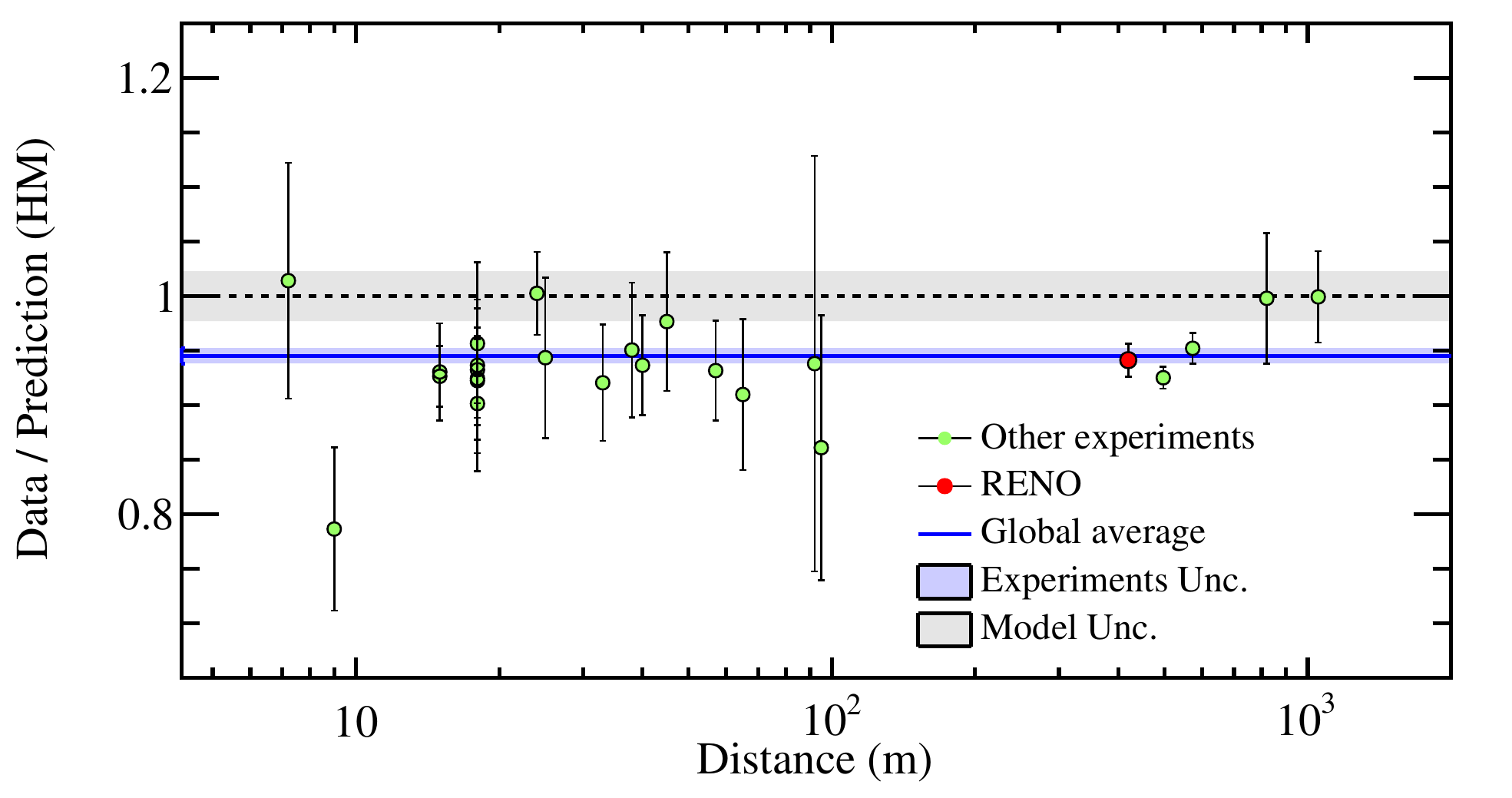}
\caption{Measured reactor $\overline{\nu}_e$ rates as a function of the distance from a reactor, relative to the HM prediction. The shaded band around unity represents the model uncertainty. The measured rate is corrected for the three flavor neutrino oscillations at each baseline\,\cite{Daya_absolute,DoubleChooz2020Nature,Dayabay_flux_2018PRD,RAA2011,Nucifer}.}
\label{fig:flux_global}
\end{figure}

\par The predicted IBD yield per fission of the $i$-th isotope is obtained as $y_i=\int \sigma(E_\nu)\phi_i(E_\nu)dE_\nu$ where the cross section of the IBD reaction, $\sigma(E_\nu)$, is used in the Ref.\,\cite{vogel99}, the input neutron live time is (879.6\,$\pm$\,0.8)\,s\,\cite{pdg2020}, and $\phi_i(E_\nu)$ is a $\overline{\nu}_e$ reference energy spectrum of the $i$-th isotope\,\cite{Huber2011,Mueller2011}. Based on the measured $R$, the IBD yield of $\overline{y}_f$ is obtained to be [5.852\,$\pm$\,0.006\,(stat.)\,$\pm$\,0.094(sys.)]\,$\times 10^{-43}$\,cm$^2$/fission.
	
\par A reactor $\overline{\nu}_e$ spectrum can be obtained by unfolding the effects of detector resolution and neutrino interaction from a measured IBD prompt spectrum. Fig.~\ref{fig:Near_prompt} shows an observed prompt energy spectrum based on 966\,094\, IBD candidate events in the near detector. A spectrum-only comparison is made by normalizing the HM prediction to the observed rate outside the prompt energy range of $3.8<E_p<6.7$\,MeV where a reasonable consistency between the data and the HM prediction is seen. A scaled HM* is defined by 0.914\,HM for the normalization. The spectral ratio between the data and the prediction shows a clear excess of observed IBD events near 5\,MeV.  A strong correlation is observed between the 5\,MeV excess, and the reactor thermal power, indicating the excess associated with the reactor\,\cite{RENO2018fueldependent}. The 5\,MeV excess was first reported by the RENO collaboration in 2014\,\cite{RENO_5MeV_boston} and other experiments\,\cite{DayaBay2016_5MeV,DoubleChooz2020Nature,NEOS2016prl,Goesgen} as well. The excess is also seen by the experiments using reactors highly enriched in $^{235}U$\,\cite{PROSPECT2020,STEREO2020}. 
	
\begin{figure}[t!]
\includegraphics[width=0.47\textwidth]{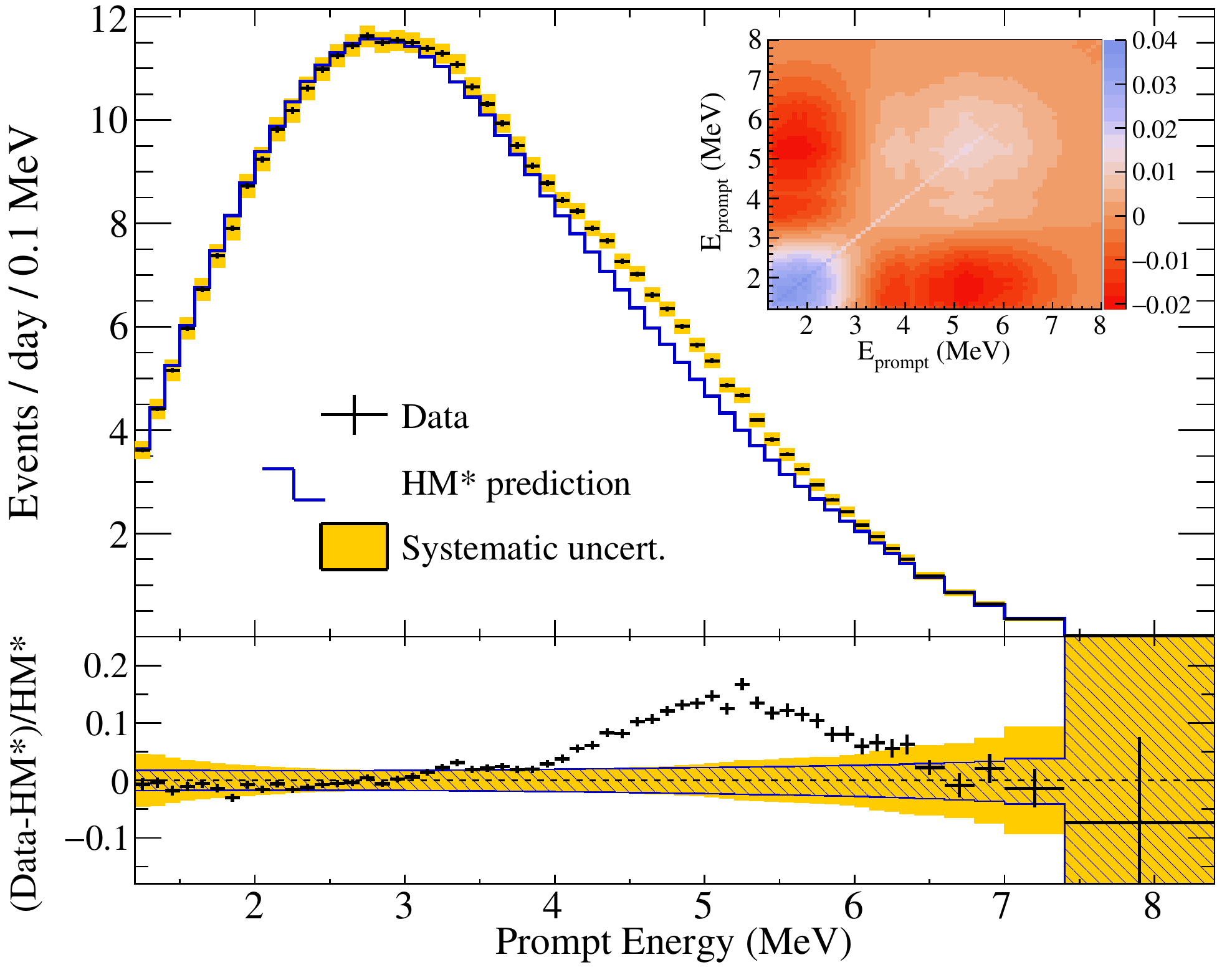}
\caption{Top: Spectral shape comparison of the observed IBD prompt energy spectrum (cross) in the near detector after the background subtraction and the scaled HM prediction (histogram). The two spectra are normalized in the energy region outside $3.8<E_p<6.7$\,MeV. The systematic uncertainty as a function of prompt energy is shown by the elements of a covariance matrix in the inset. The energy-dependent uncertainties only are shown in the inset where the uncertainty of detection efficiency is not included. Bottom: Spectral ratio between the observed spectrum and the scaled HM prediction. The error bars represent statistical errors. The yellow band corresponds to the total systematic uncertainty, the magnitude of the diagonal elements in the complete covariance matrix. The blue shaded band represents the uncertainty of the scaled HM prediction including the reactor-related uncertainties.}
\label{fig:Near_prompt}
\end{figure}
			
\begin{figure}[t!]
\includegraphics[width=0.48\textwidth]{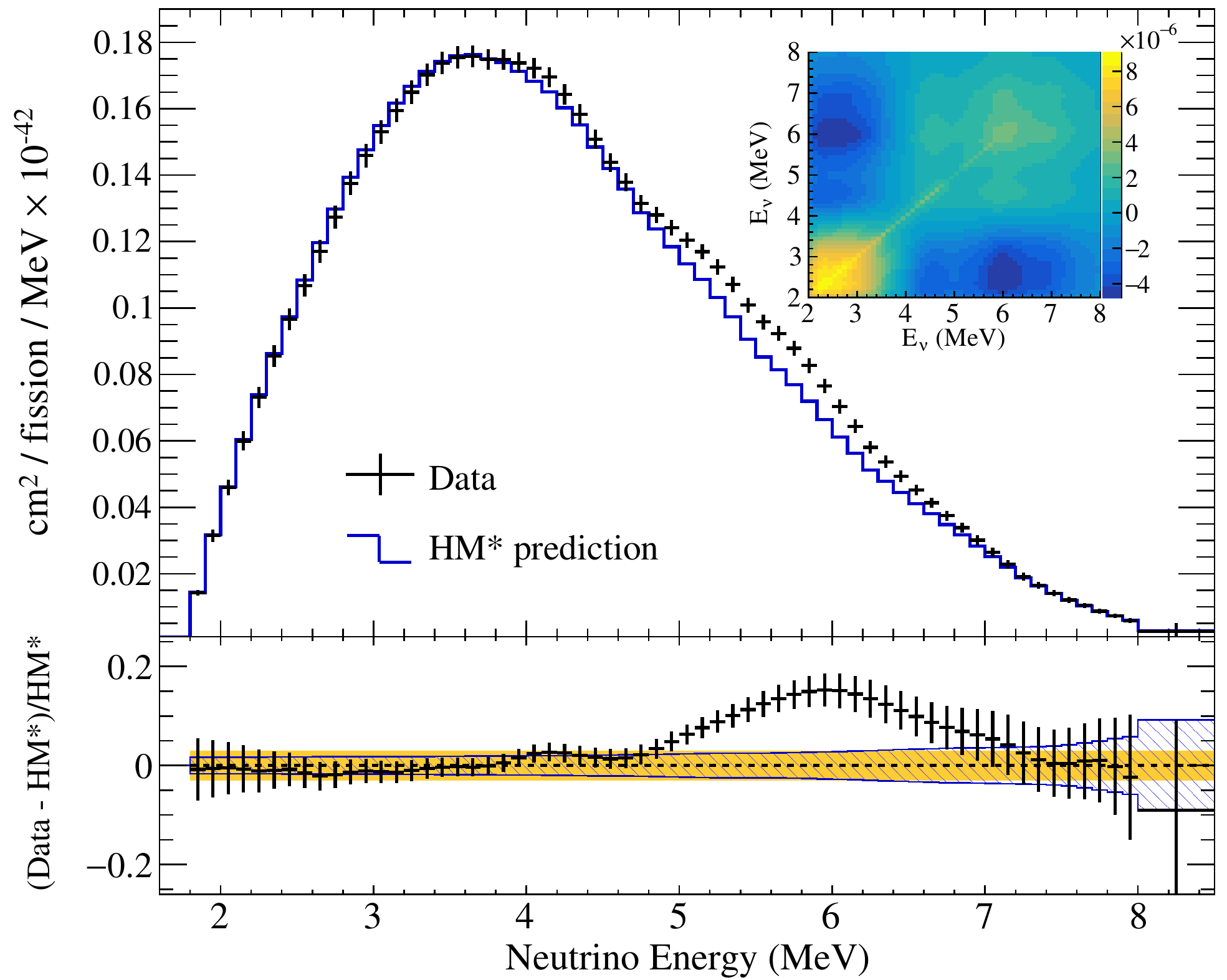}
\caption{Top: The obtained reactor $\overline{\nu}_e$ spectrum (cross) from unfolding and the scaled HM prediction (histogram) for comparison. The oscillation effect is removed using the measured $\theta_{13}$ to obtain the spectrum at the reactor. The two spectra are normalized outside the 6\,MeV excess region of $4.6<E_{\nu}<7.4$\,MeV. The data error bar represents the total uncertainty including the statistical and systematic errors. The covariance matrix obtained from unfolding is shown in the inset. Bottom: Ratio of the extracted $\overline{\nu}_e$ spectrum to the scaled HM prediction. The blue shaded band shows the uncertainty of the scaled HM prediction including the reactor-related uncertainties. The uncertainty of the summation model predictions is also shown as a yellow band.}
\label{fig:absolute_flux}
\end{figure}
		
\par The observed IBD prompt spectrum contains several detector response effects including conversion of the neutrino energy to the prompt energy, prompt energy resolution, nonlinearity of energy scale, and energy loss in the acrylic vessel. An accurate prompt-energy measurement is crucial for extracting the reactor $\overline{\nu}_e$ spectrum. The energy scale is calibrated using several radioactive sources and neutron capture events. The energy scale uncertainty is largely attributed to the nonlinear response of scintillating energy, mainly due to the quenching effect and Cherenkov radiation. A more detailed description of the energy nonlinearity is given in Ref.\,\cite{RENO2018PRD}. The energy resolution is roughly 7\% at 1\,MeV and 3\% at 7\,MeV\,\cite{RENO2018PRD}. These detector response effects are simulated as closely as possible in the IBD MC sample. A simulated prompt energy spectrum is used as a training sample to unfold the detector response effects from the observed spectrum. 

\par The uncertainty in the unfolded spectrum arises from the prompt  spectrum, associated with an imperfect understanding of the detector response effects in the simulation. The uncertainty is evaluated using a large number of modified HM prompt energy spectra that are generated with the 5MeV excess within the detector response uncertainties. A covariance matrix, consisting of energy correlated and uncorrelated components, is constructed from energy-dependent uncertainties as shown in the inset of Fig.~\ref{fig:Near_prompt}. A major uncertainty comes from the energy scale uncertainty and is estimated by a toy MC sample using varied charge-to-energy conversion functions within its uncertainty. The uncertainty is either correlated or anti-correlated among the energy bins and the uncertainty size is estimated to be 6\% at 1\,MeV, 0.4\% at 3\,MeV, and 7\% at 7\,MeV. The background and spill-in uncertainties also contribute to the energy-dependent uncertainty to the unfolding. A dominant source of uncertainty below 1\,MeV is the spill-in rate uncertainty associated with the energy loss in the acrylic vessel. The energy-uncorrelated uncertainties come from the background spectrum and statistical uncertainties. The energy-independent uncertainties of detection efficiency and reactors are not considered in the unfolding process but included as additional uncertainties to the unfolded spectrum.

\begin{figure}[t!]
\includegraphics[width=0.495\textwidth]{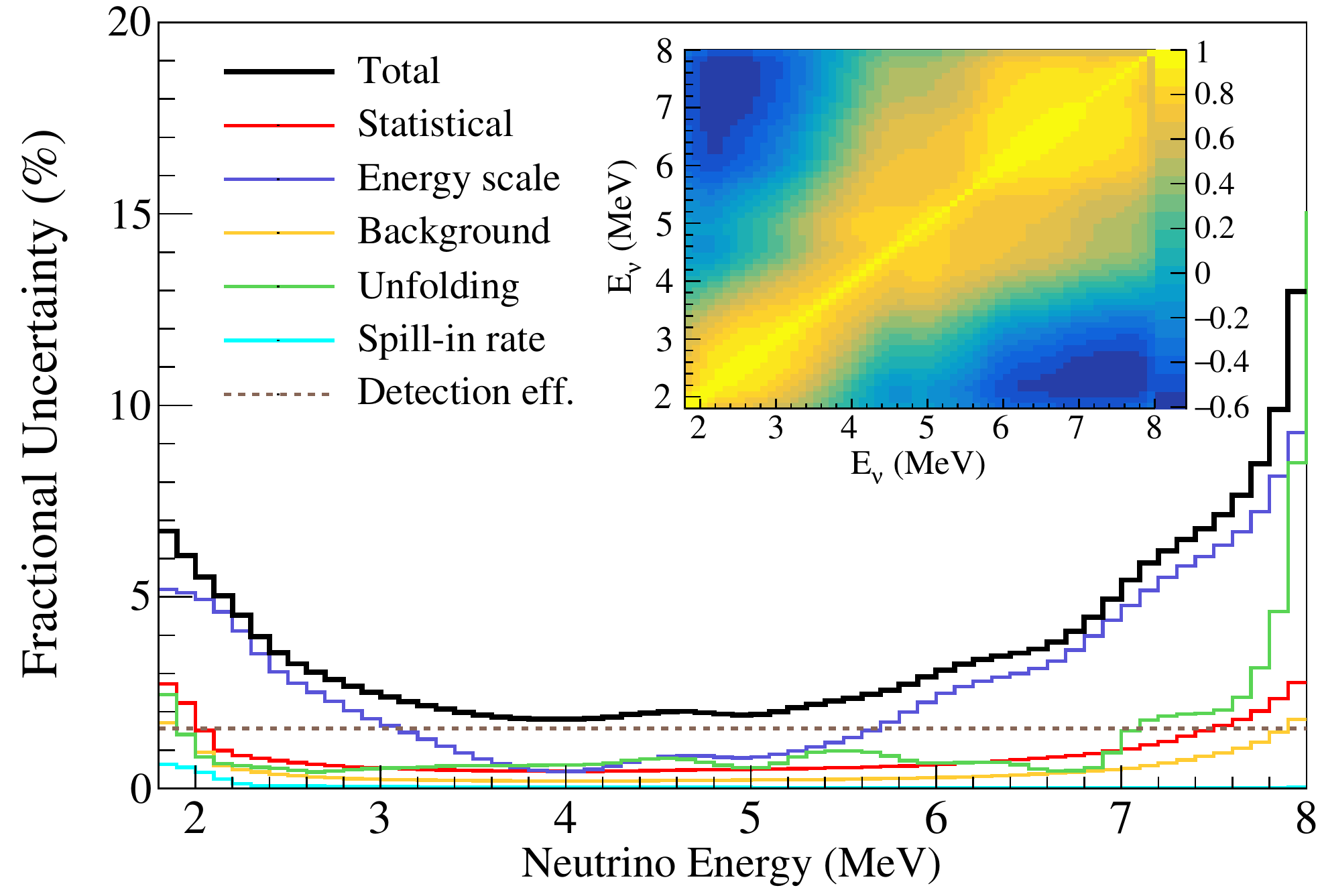}
\caption{Uncertainties of the obtained reactor $\overline{\nu}_e$ spectrum. The correlation matrix of the $\overline{\nu}_e$ spectrum is shown in the inset.}
\label{fig:fractional_uncert}
\end{figure}

\par The detector response effects are removed by finding a true $\overline{\nu}_e$ energy relating to observed prompt energy using a training sample. Because of finite statistics in an observed spectrum, the unfolding process cannot find an exact one-to-one correspondence between them. To mitigate the ill-posed problem, proper regularization is adopted by two unfolding methods of Iterative Bayesian Unfolding (IBU)\,\cite{IBU} and Singular Value Decomposition (SVD)\,\cite{SVD}. The IBU algorithm applies an iterative Bayesian approach to update an initial prior spectrum by a feedback process while the SVD algorithm implements the initial spectrum as a constraint. Both algorithms give consistent results and their difference averaged in energy is 0.14\%. In this analysis, the IBU algorithm is chosen and the difference is taken as a systematic uncertainty of an unfolded result.

\par The unfolding algorithm, together with the uncertainties of the observed prompt spectrum, is implemented in the ROOT unfolding framework (RooUnfold)\,\cite{RooUnfold}. A number of iterations regulate weighting for an initial prior spectrum and an observed spectrum. The weighting balance is chosen based on the L-curve\,\cite{Lcurve1995}. The fourth iteration is found to produce the best solution for the RENO data. The systematic uncertainty associated with the iteration is estimated by varying the number of iterations and 0.15\% as an average in energy. The uncertainty due to the MC statistical fluctuation is estimated to be 0.07\%. The energy resolution uncertainty contributes 0.03\% as an average in energy and 0.22\% at $E_\nu = 1.8$\,MeV due to the prompt energy threshold.

\par  By using the HM model for the initial prior spectrum a smooth function constraint is implicitly assumed by the unfolding. This assumption may ignore possible substructures in the unfolded $\overline{\nu}_e$ spectrum. Recently, a possible fine structure in the reactor $\overline{\nu}_e$ spectrum is discussed seriously\,\cite{Danielson2018,Forero2017} for a future high resolution detector \,\cite{JUNO2015}. Two summation models including a fine spectral structure\,\cite{Sonzongni2018,abinitio2018} are used for the initial prior spectra and compared to the unfolding result of the HM model. The model dependence on the initial prior spectrum is estimated to be 0.64\%. The energy spectrum uncertainties of both HM and summation models, roughly 3\%, introduce an additional unfolding error as the ambiguity of initial prior spectrum. 

\par The total uncertainty of the unfolding process comes from MC training, iteration, algorithm dependence, and initial prior spectrum dependence and is estimated to be 0.69\% as an average in energy. The covariance matrix of the unfolded reactor $\overline{\nu}_e$ spectrum is obtained according to the error propagation from the observed prompt spectrum\,\cite{IBU,RooUnfold}.

\par The obtained reactor $\overline{\nu}_e$ spectrum and its covariance matrix in the $\overline{\nu}_e$ energy are shown in the top panel of Fig.~\ref{fig:absolute_flux}. The oscillation effect is removed using the best-fit result of $\theta_{13}$ to obtain the $\overline{\nu}_e$ spectrum at the source. The bottom panel of Fig.~\ref{fig:absolute_flux} presents the ratio of the extracted reactor $\overline{\nu}_e$ spectrum to the scaled HM prediction. The extracted spectrum shows a clear excess now near 6\,MeV relative to the HM prediction. The systematic errors and their correlation matrix elements are shown in Fig.~\ref{fig:fractional_uncert}. The correlation matrix also includes detection efficiency and reactor-related errors. The total error averaged in energy, including detection efficiency, is 2.6\% and the second largest contribution of 1.59\% comes from the energy scale uncertainty. The errors of statistical fluctuation, backgrounds, unfolding, spill-in, and detection efficiency are 0.58\%, 0.27\%, 0.69\%, 0.04\%, and 1.56\%, respectively. Additional precision measurements are needed to understand the origin of the 6\,MeV excess. The obtained reactor $\overline{\nu}_e$ spectrum with the detector response effects unfolded can be directly compared or combined with other measured spectra for studying unknown neutrino properties and reactor models\,\cite{NEOS2016prl,renoneossterile,jointDayaBayProspect}. The prompt IBD spectrum, the covariance matrix of the prompt spectrum, the antineutrino spectrum, the covariant matrix of the antineutrino spectrum, and the detector response matrix are reported as a supplementary material of this Letter\,\cite{RENO2021DataRelease}.

\par In summary, RENO's first results on the flux and energy spectrum of reactor $\overline{\nu}_e$ are obtained from roughly 1\,million IBD candidate events. The observed IBD yield is measured as (5.852$\,\pm\,$0.094$) \times 10^{-43}$\,cm$^2$/fission, corresponding to 0.941\,$\pm$\,0.015 of the HM prediction. This confirms the deficit of observed reactor $\overline{\nu}_e$ rate reassuring the reactor antineutrino anomaly. A reactor $\overline{\nu}_e$ spectrum is obtained by removing both detector response and $\theta_{13}$ oscillation effects from the measured IBD prompt spectrum. The $\overline{\nu}_e$ spectrum shows a clear excess around 6\,MeV relative to the HM prediction. The discrepancies of reactor $\overline{\nu}_e$ flux and spectrum between this measurement and the prediction suggest modification and reevaluation of the current reactor $\overline{\nu}_e$ models. The next round of precision measurements will provide useful information on the origin of the 6\,MeV excess.\\

\par The RENO experiment is supported by the National Research Foundation of Korea (NRF) grants No. 2009-0083526, No. 2019R1A2C3004955, and 2017R1A2B4011200 funded by the Korea Ministry of Science and ICT. Some of us have been supported by a fund from the BK21 of NRF. This work was partially supported by the New Faculty Startup Fund from Seoul National University. We gratefully acknowledge the cooperation of the Hanbit Nuclear Power Site and the Korea Hydro \& Nuclear Power Co., Ltd. (KHNP). We thank KISTI for providing computing and network resources through GSDC, and all the technical and administrative people who greatly helped in making this experiment possible. \\
	
\bibliography{reno_absolute}
	
\end{document}